\documentclass[twocolumn,showpacs,preprintnumbers,amsmath,amssymb,superscriptaddress,prl]{revtex4-1}
\usepackage{epsfig}
\usepackage{graphicx}
\usepackage{color}
\usepackage{bm}

\begin{document}

\title{Current induced torques in structures with ultra-thin IrMn antiferromagnet}

\author{H.~Reichlov\'a}
\affiliation{Institute of Physics ASCR, v.v.i., Cukrovarnick\'a 10, 162 53
Praha 6, Czech Republic}
\affiliation{Faculty of Mathematics and Physics, Charles University in Prague,
Ke Karlovu 3, 121 16 Prague 2, Czech Republic}
\author{D.~Kriegner}
\affiliation{Faculty of Mathematics and Physics, Charles University in Prague,
Ke Karlovu 3, 121 16 Prague 2, Czech Republic}
\author{V.~Hol\'y}
\affiliation{Faculty of Mathematics and Physics, Charles University in Prague,
Ke Karlovu 3, 121 16 Prague 2, Czech Republic}
\author{K.~Olejn\'{\i}k}
\affiliation{Institute of Physics ASCR, v.v.i., Cukrovarnick\'a 10, 162 53
Praha 6, Czech Republic} 
\author{V.~Nov\'ak}
\affiliation{Institute of Physics ASCR, v.v.i., Cukrovarnick\'a 10, 162 53
Praha 6, Czech Republic} 
\author{M.~Yamada}
\affiliation{Hitachi Ltd., Central Research Laboratory, 1-280 Higashi-koigakubo, Kokubunju-shi, Tokyo 185-8601, Japan} 
\author{K.~Miura}
\affiliation{Hitachi Ltd., Central Research Laboratory, 1-280 Higashi-koigakubo, Kokubunju-shi, Tokyo 185-8601, Japan} 
\author{S.~Ogawa}
\affiliation{Hitachi Ltd., Central Research Laboratory, 1-280 Higashi-koigakubo, Kokubunju-shi, Tokyo 185-8601, Japan} 
\author{H.~Takahashi}
\affiliation{Hitachi Ltd., Central Research Laboratory, 1-280 Higashi-koigakubo, Kokubunju-shi, Tokyo 185-8601, Japan} 
\author{T.~Jungwirth}
\affiliation{Institute of Physics ASCR, v.v.i., Cukrovarnick\'a 10, 162 53
Praha 6, Czech Republic} 
\affiliation{School of Physics and
Astronomy, University of Nottingham, Nottingham NG7 2RD, United Kingdom}
\author{J.~Wunderlich}
\affiliation{Institute of Physics ASCR, v.v.i., Cukrovarnick\'a 10, 162 53 Praha 6, Czech Republic}
\affiliation{Hitachi Cambridge Laboratory, Cambridge CB3 0HE, United Kingdom}

\begin{abstract}
Relativistic current induced torques and devices utilizing antiferromagnets have been independently considered as two promising new directions in spintronics research. Here we report electrical measurements of the  torques in structures comprising a $\sim1$~nm thick layer of an antiferromagnet IrMn. We perform ac-electrical measurements in Ta/IrMn/CoFeB and Ta/IrMn structures below and above the N\'eel temperature of the ultra-thin IrMn and compare the results with control experiments in a Ta/CoFeB paramagnet/ferromagnet bilayer. The torques observed in CoFeB  are consistent with the opposite sign  of the spin Hall angle in Ta and IrMn. By combining temperature-dependent measurements in all three structures we conclude that torques act also in the antiferromagnetic IrMn at the interface with Ta. 
\end{abstract}


\maketitle
Since the relativistic spin-orbit interaction couples electron's momentum and spin it can lead to a range of effects when systems are brought out of equilibrium by applied electric fields. Non-equilibrium spin polarization phenomena may occur even in non-magnetic spin-orbit coupled conductors. A prime example is the spin Hall effect (SHE) in which an electrical current passing through a material with relativistic spin-orbit coupling generates a transverse spin-current polarized perpendicular to the plane defined by the charge and spin-current \cite{Sinova2014}. The spin-current generated in a normal metal (NM) via the SHE can be absorbed in a ferromagnet (FM) and the spin angular momentum transferred to the magnetization. The corresponding relativistic spin torques are a candidate spintronic technology for a new generation of electrically-controlled spintronic devices \cite{Miron2011b,Liu2012}.

Unlike FMs, antiferromagnets (AFMs) have traditionally played only a static supporting role in spintronic devices by enhancing the magnetic hardness of the FM electrode via the interfacial exchange-bias effect \cite{Radu2008}. More recently, AFM spintronic devices have been demonstrated, including magneto-resistors and memories \cite{Park2011b,Wang2012a,Marti2014,Fina2014}, in which AFM electrodes play an active role in the device. 

A key challenge in the field of AFM spintronics is to find efficient means for manipulating the compensated moments. Here the relativistic spin torques represent one of the promising routes \cite{Zelezny2014}. In one picture, the torques occur in AFMs like Mn$_2$Au or CuMnAs whose spin-sublattices of the bulk crystal structure form inversion partners \cite{Zelezny2014,Wadley2015}. 

In this paper we focus on an alternative scenario applying to AFMs of a general bulk crystal structure that are prepared in thin films and interfaced with a FM and/or a spin-orbit coupled NM. The SHE generated in the AFM \cite{Mendes2014,Zhang2014e} can induce a torque on the adjacent FM \cite{Tshitoyan2015}.  Remarkably, theory also predicts \cite{Zelezny2014} that relativistic current-induced torques at the NM/AFM interface might provide an efficient mechanism for manipulating the AFM spins. 

Our aim is to experimentally identify these phenomena  by performing systematic ac-electrical measurements in a NM/AFM/FM trilayer, NM/AFM bilayer, and a control NM/FM bilayer, with the structures comprising  Ta NM, IrMn AFM, and CoFeB FM films. The method we employ is based on utilizing  ultra-thin ($\sim 1$~nm) Ir$_{0.2}$Mn$_{0.8}$.  The small thickness of the AFM film is favorable for detecting spin-torques generated  at the NM/AFM interface. Moreover, the N\'eel temperature  is suppressed from the bulk $T_N=700$~K \cite{Yamaoka1974} to well below room-temperature \cite{Petti2013} in the ultra-thin IrMn. This allows us  to compare the observed torques in both the NM/AFM and NM/AFM/FM stacks when IrMn is in the  antiferromagnetic and paramagnetic phases and  to confirm the magnetic origin of the current-induced effects observed in these structures. 
 
We start with discussing our experiments in the NM/AFM/FM trilayer. The layers were deposited by UHV RF magnetron sputtering on a thermally oxidized Si substrate (700~nm SiO$_2$ on (001) Si) at a base pressure of 10$^{-9}$~Torr and in a magnetic field of 5~mT. To achieve high crystal and magnetic quality of the films, the wafers were annealed at 350$^\circ$C for 1~hour in a 10$^{-6}$~Torr vacuum in a magnetic field of 0.6~T applied along the sample edge \cite{Hayakawa2005}. (For more details see the Supplementary information  \cite{suppl}.) From simulations of the X-ray reflectivity data \cite{suppl} we obtain the following structure: Ta($2.2\pm0.2$)/IrMn($0.6\pm0.3$)/CoFeB($0.9\pm0.2$)/MgO($1.4\pm0.2$) capped with 10~nm of Al$_2$O$_3$. The numbers in brackets correspond to the layer thicknesses in nm and are consistent with the nominal growth values.  

\begin{figure}[h]
\hspace*{0cm}\epsfig{width=1\columnwidth,angle=0,file=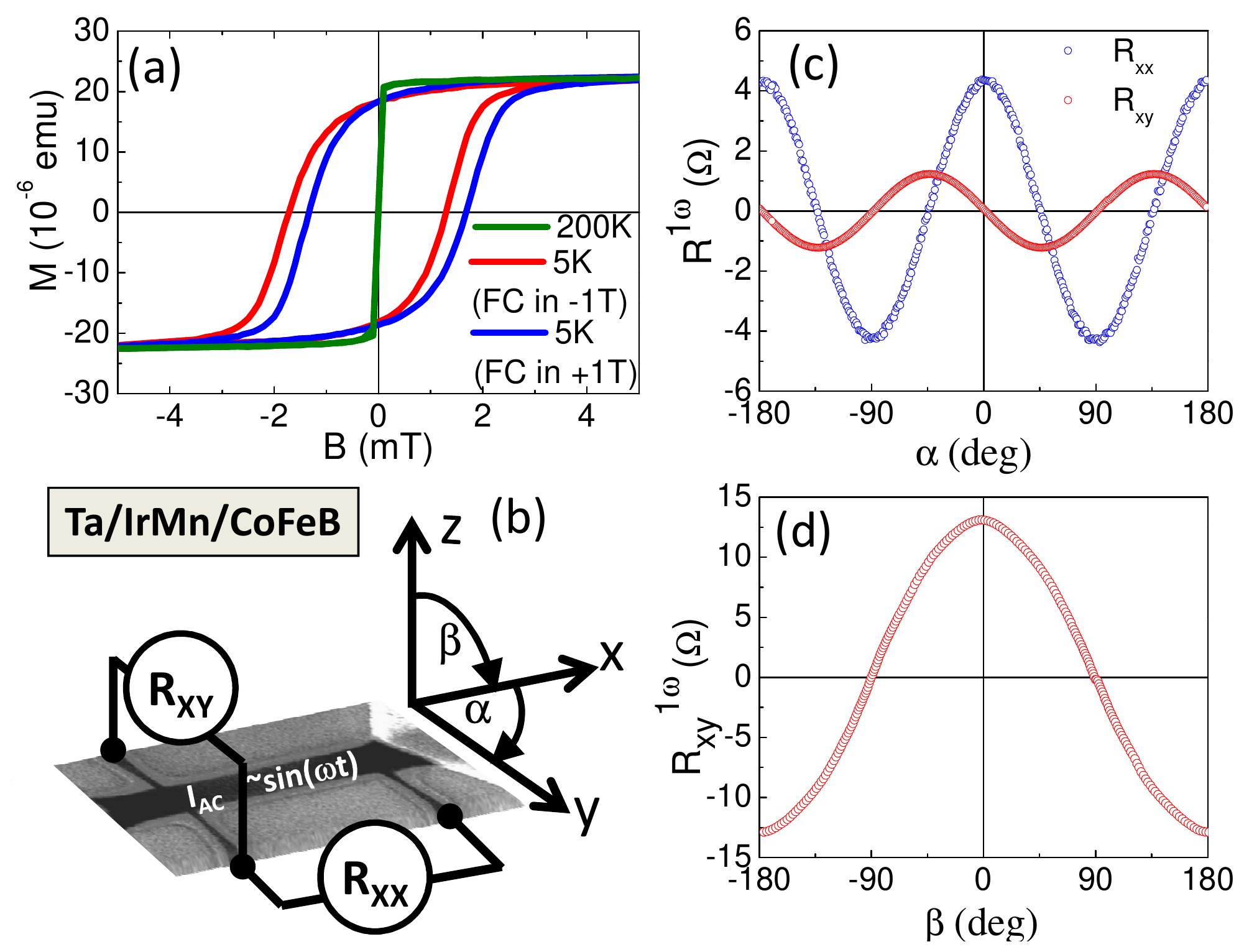}
%
\caption{
(a) SQUID magnetization measurement after  cooling the sample from 400~K to 5~K in magnetic field +1~T (blue line) and -1~T (red line), and after field cooling to 200~K (green line). (b) Electron microscope image of the Hall bar device and a schematic of our experimental setup. (c) Variations of $R_{xx}^{1\omega}$ and $R_{xy}^{1\omega}$ when rotating magnetization in plane. The average $R_{xx}^{1\omega}$ is 1.7~k$\Omega$. (d) $R_{xy}^{1\omega}$ for out-of-plane rotation. All data are for the Ta/IrMn/CoFeB trilayer.}
\label{Rf1}
\end{figure}

Magnetic measurements shown in Fig.~\ref{Rf1}(a) were performed by the superconducting quantum interference device (SQUID). Exchange coupling at low temperatures between the IrMn AFM and the CoFeB FM is confirmed by the shift and broadening of the hysteresis loop when cooling the sample from 400~K to 5~K in a magnetic field of $\pm1$~T applied along the field direction used during  annealing. Note that  the low-temperature exchange bias is weak and positive, consistent with the small thickness of our IrMn film and with previous studies of  IrMn/CoFeB interfaces \cite{Raju2013}. Consistent with the expected high-temperature paramagnetic phase of our ultra-thin IrMn, the sample shows no shift and broadening of the hysteresis loop when field-cooled to 200~K. 

The device geometry is shown in Fig.~\ref{Rf1}(b). The Hall bar width is 2~$\mu$m and the distance between longitudinal contacts is 7~$\mu$m  \cite{suppl}. In electrical measurements, the Hall bar is biased by a low frequency (123 Hz) ac current $I_{AC}=I_0\sin(\omega t)$ applied along the x-axis. We use lock-in amplifiers to measure simultaneously first harmonic (1$\omega$) and second harmonic (2$\omega$) signals \cite{Pi2010,Garello2013,Kim2013}.  Examples of the first harmonic signals  are shown in Figs.\ref{Rf1}(c),(d) where angles $\alpha$ and $\beta$ correspond to in-plane ($x-y$) and out-of-plane ($x-z$) magnetization rotation, respectively. The relative amplitudes of the longitudinal and transverse anisotropic magnetoresistance (AMR) seen in Fig.\ref{Rf1}(c)  are consistent with the aspect ratio of the longitudinal and transverse dimensions of the Hall bar. When rotating the magnetization out-of-plane, the $R_{xy}^{1\omega}$ signal is dominated by the anomalous Hall effect (AHE), as seen in Fig.\ref{Rf1}(d). 

Current induced torques contribute to the higher-harmonic non-ohmic components of the resistance. In our study we focus on the second harmonic signals. For a systematic set of measured first and second harmonic transport signals, including also experiments for magnetization rotation in the $y-z$ plane and magnetic-sweep measurements, we refer to the Supplementary information \cite{suppl}. 

\begin{figure}[h]
\hspace*{0cm}\epsfig{width=1\columnwidth,angle=0,file=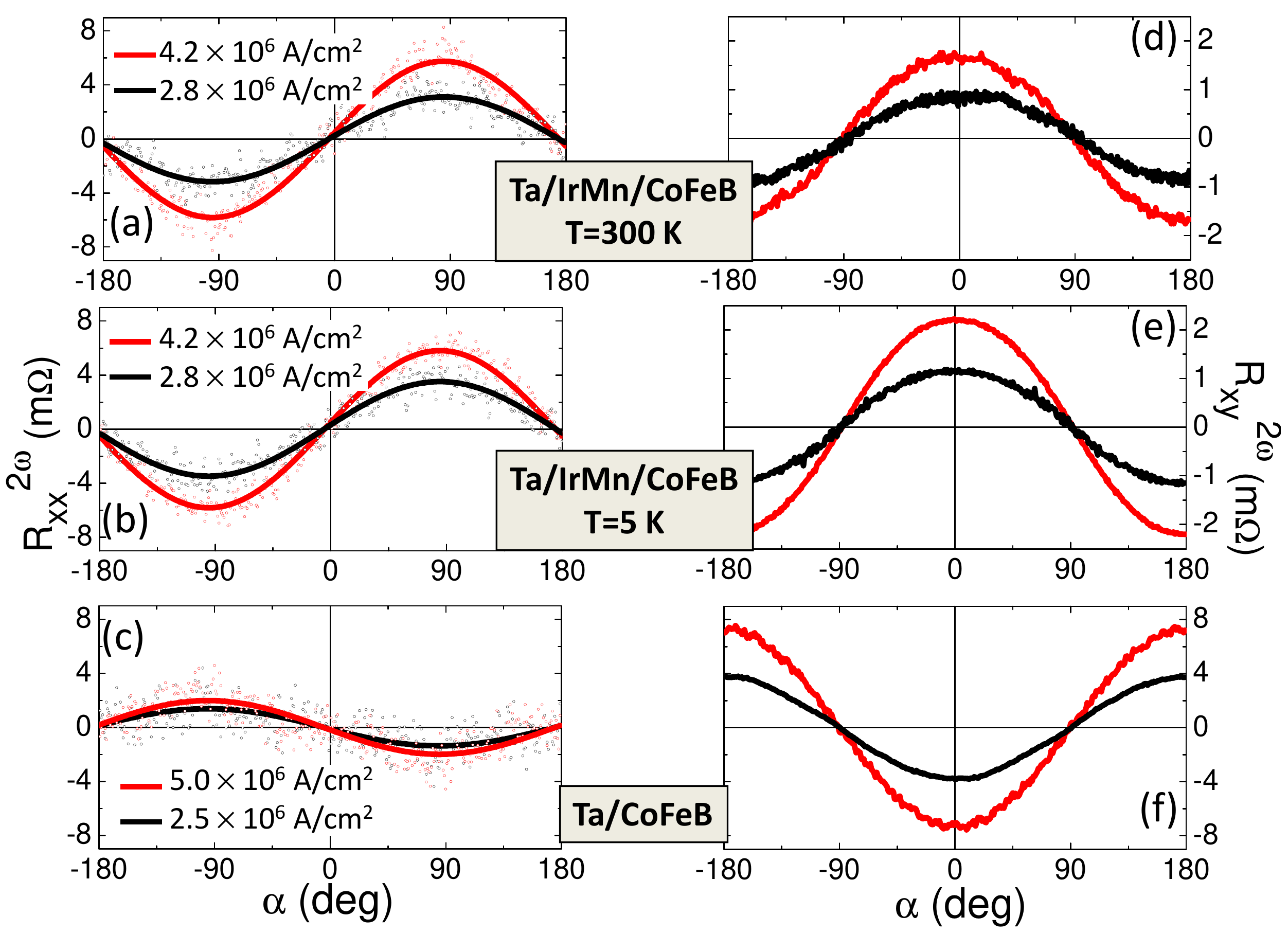}
%
\caption{Angular dependence of the second harmonic signals $R_{xx}^{2\omega}$ (a)-(c) and $R_{xy}^{2\omega}$ (d)-(f) measured in a saturating in-plane magnetic field in the Ta/IrMn/CoFeB samples at 300~K and 5~K, and in the reference Ta/CoFeB sample. Data measured at two current densities are shown in each panel.}
\label{Rf2}
\end{figure}

Examples of $R_{xx}^{2\omega}$ and $R_{xy}^{2\omega}$ for the in-plane rotation of magnetization are shown in Fig.~\ref{Rf2}. Two types of torques can be expected in our structure: The antidamping-like torque  which for magnetization ${\bf m}$ rotating in the plane is driven by an out-of-plane effective field $\sim {\bf m}\times{\bf \hat{y}}$, and the field-like torque  due to an in-plane field $\sim {\bf \hat{y}}$. Neither of these fields would generate via the AMR a signal $R_{xx}^{2\omega}\sim\sin\alpha$ (see Tab.~S1 in the Supplementary information \cite{suppl}) that is seen in  Figs.~\ref{Rf2} (a)-(c) for the Ta/IrMn/CoFeB sample at high and low temperatures and also for the control Ta/CoFeB sample. 

A mechanism that can consistently explain the angular dependence of $R_{xx}^{2\omega}$ seen in Figs.~\ref{Rf2}(a)-(c) is the longitudinal spin Seebeck effect (SSE) \cite{suppl}. The SSE is generated by an out-of-plane thermal gradient. It drives a spin current from the FM into the adjacent layer and here the inverse SHE converts the spin current into an electrical voltage along the in-plane direction perpendicular to the magnetization in the FM \cite{Kikkawa2013b}. While not detecting the spin torque itself in $R_{xx}^{2\omega}$, the data provide us, nevertheless,  with a valuable information on the SHE which can generate the current-induced spin torque. Signatures of the spin torque will be identified below in the discussion of the $R_{xy}^{2\omega}$ signals. 

Earlier studies found opposite sign of spin Hall angles in IrMn and Ta   \cite{Liu2012,Mendes2014,Zhang2014e} which is consistent with the measured opposite sign of the $R_{xx}^{2\omega}$ in the Ta/IrMn/CoFeB and the Ta/CoFeB samples and confirms that these longitudinal signals can originate from the SSE. Thermal effects generated by Joule heating are quadratic in the driving current and can, therefore, contribute to the second harmonic signals for the biasing current $I_{AC}=I_0\sin(\omega t)$. Consistent with the data in Figs.~\ref{Rf2}(a)-(c), the signals due to the SSE scale with the current amplitude since the thermally generated voltage is quadratic in $I_0$ and the resistances plotted in  Figs.~\ref{Rf2}(a)-(c) are obtained by dividing the voltages by $I_0$.  

The observed SSE signals are independent of the base temperature. In the transport experiments, the vertical thermal gradient across the IrMn/CoFeB bilayer (or the control Ta/CoFeB bilayer) is  determined by the gradient in the resistivity across the bilayer. Since the dependence of the resistivity on base temperature is negligible in the studied temperature range (see Fig.~S3 in the Supplementary information \cite{suppl}), the Joule heating gradient across the metal films can be also expected to  be independent of the base temperature. From this we conclude, that the inverse SHE in IrMn does not change significantly between the low-temperature AFM phase and the high-temperature paramagnetic phase. 

The SSE contributes to the transverse resistance as $R_{xy}^{2\omega}\sim\cos\alpha$ (see Tab.~S2 in the Supplementary information \cite{suppl}). This angular dependence is observed in Figs.~\ref{Rf2}(d)-(f). However, when comparing in Figs.~\ref{Rf2}(a) and (d) the amplitudes of $R_{xx}^{2\omega}$ and $R_{xy}^{2\omega}$ signals for the Ta/CoFeB sample they do not scale with the Hall bar aspect ratio. As in a number of previous studies of NM/FM bilayers, the additional $R_{xy}^{2\omega}$ signal is attributed to the current-induced antidamping-like torque acting in CoFeB.  The corresponding out-of-plane effective field $\sim {\bf m}\times{\bf \hat{y}}$ can originate form the spin-current polarized along the y-axis which is generated by the SHE in Ta and  absorbed in the CoFeB FM. The $R_{xy}^{2\omega}\sim\cos\alpha$ symmetry reflects the ouf-of-plane tilt of the magnetization due to the current-induced effective out-of-plane field which is sensed electrically via the AHE in CoFeB \cite{suppl}. 

In the Ta/IrMn/CoFeB sample the signal due to the antidamping-like torque is detected at low temperatures. In the high-temperature paramagnetic phase of IrMn the amplitudes of $\alpha$-dependent $R_{xx}^{2\omega}$ and $R_{xy}^{2\omega}$ signals scale with the Hall bar aspect ratio, implying that the SSE dominates and the current induced torque is diminished. 

\begin{figure}[h]
\hspace*{0cm}\epsfig{width=1\columnwidth,angle=0,file=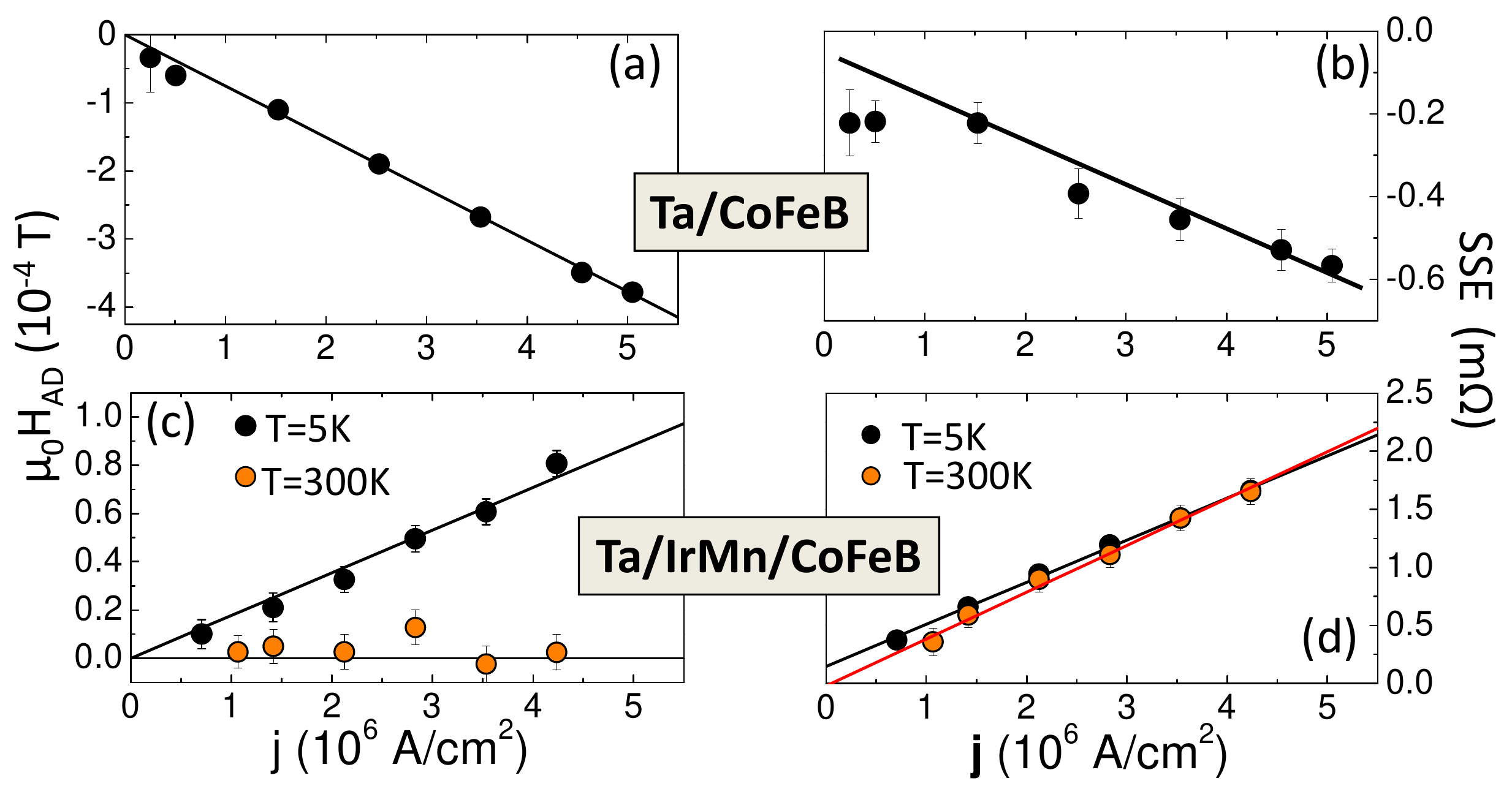}
%
\caption{(a) Current density dependence of the current-induced effective out-of-plane field $H_{AD}$ (black dots) in the reference Ta/CoFeB sample.  (b) Current density dependence of the SSE contribution recalculated for one square. (c,d) Same as (a,b) for the Ta/IrMn/CoFeB sample at 5~K and 300~K.}
\label{f3}
\end{figure}

In Fig.~\ref{f3} we summarize the low and high-temperature measurements in the Ta/IrMn/CoFeB sample and compare with the temperature-independent results in the control Ta/CoFeB sample. After subtracting the SSE contribution, the remaining $R_{xy}^{2\omega}$ component due to the antidamping-like torque is recalculated into the magnitude $H_{AD}$ of the effective current-induced out-of-plane field. For this we  used the calibration of the first harmonic AHE signal in an externally applied out-of-plane magnetic field (for more details see Supplementary text below Tab.~S2 \cite{suppl}). Similar to the SEE signal, $H_{AD}$ scales linearly with the density of the driving ac current since the generated changes in the magnetization direction are small and AHE varies linearly in this small tilt regime. 

Fig.~\ref{f3} highlights the SHE origin of $H_{AD}$ which, as in the SSE case,  changes sign when CoFeB is interfaced with either Ta or IrMn. The inferred spin Hall angle in the Ta layer from measurements in the Ta/CoFeB bilayer is  $\approx-0.036\pm0.020$. The effective spin Hall angle recalculated to the current flowing in the IrMn layer of the Ta/IrMn/CoFeB stack at 5~K is $\approx+0.029\pm0.015$. 

We point out that the SSE and the antidamping-like spin torque, as well as the inferred magnitudes of $H_{AD}$ and the spin Hall angles, were confirmed by performing magnetization rotation experiments in all three orthogonal planes. The measured transport signals were systematically decomposed  into harmonic functions of the magnetization rotation angles. The technique is described in detail in the Supplementary material \cite{suppl}. 

We also remark that angular-dependent components with a symmetry corresponding to the field-like torque were  not identified in the measured data. One source of the field-like torque is the Oersted field which in our experiments does not exceed 0.1~mT. In data shown in Figs.~\ref{Rf2}(a)-(c), e.g.,  the corresponding $R_{xx}^{2\omega}\sim\sin2\alpha\cos\alpha$ term \cite{suppl} would contribute via AMR with an amplitude below $\sim 10^{-4}$~$\Omega$, i.e., within the noise in Figs.~\ref{Rf2}(a)-(c). 

\begin{figure}[h]
\hspace*{0cm}\epsfig{width=1\columnwidth,angle=0,file=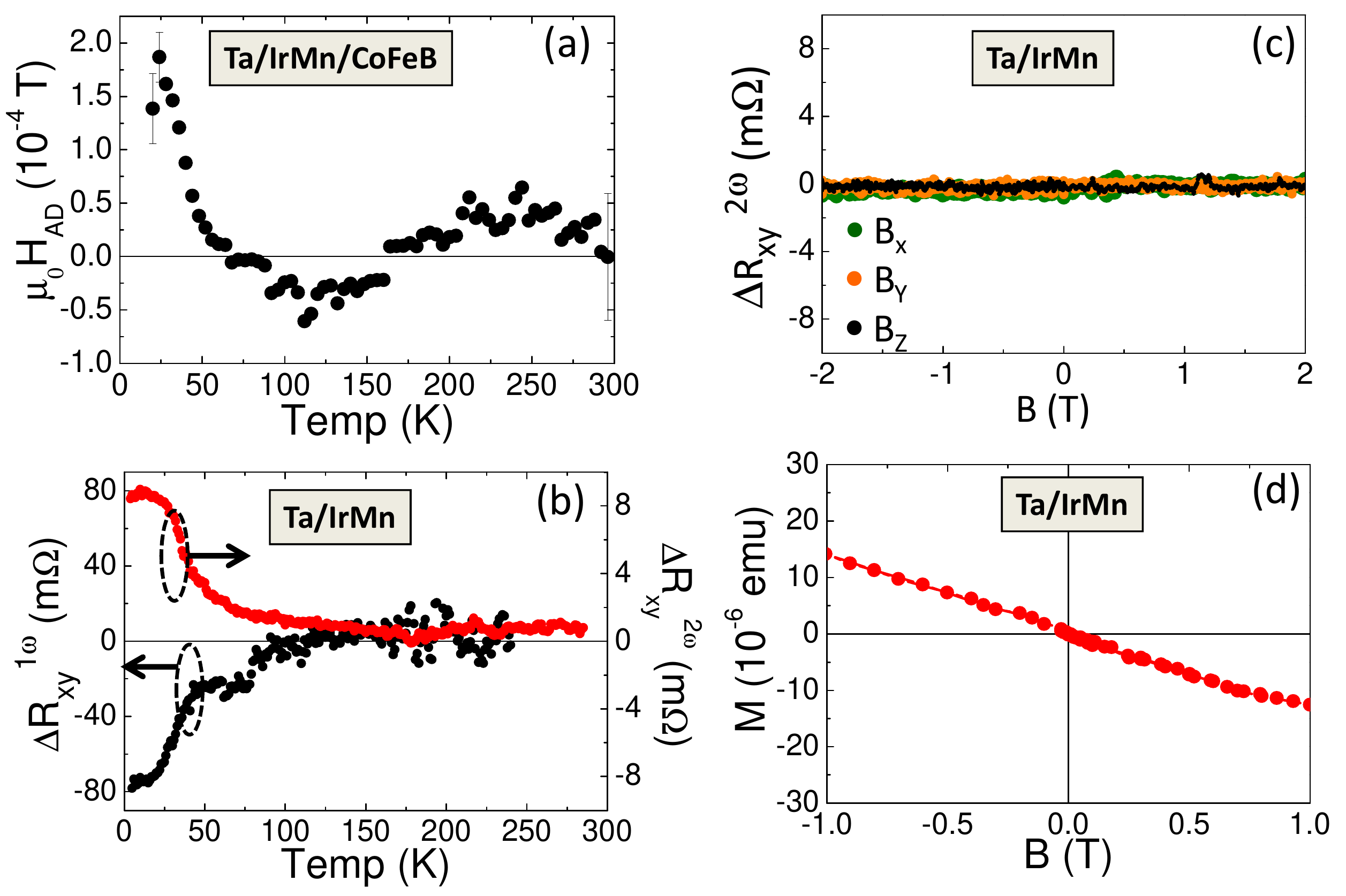}
%
\caption{(a) Temperature dependence of the current-induced effective out-of-plane field $H_{AD}$ measured in the Ta/IrMn/CoFeB sample. (b) Temperature-dependent difference in the $R_{xy}^{2\omega}$ signal in the Ta/IrMn sample measured while cooling the sample in applied in-plane (x-axis) and out-of-plane magnetic fields of 2~T. Both first harmonic (black line) and second  harmonic (red line) signals are plotted in the panel. (c) The signals become insensitive to the magnetic field when sweeping magnetic field along x, y, or y-axis at low temperature (5~K). (d) SQUID magnetization measurement of the Ta/IrMn sample showing only a diamagnetic signal from the substrate and no net moment in IrMn at low temperature (10~K).}
\label{f4}
\end{figure}

As noted above, when the spin-current is injected from the CoFeB FM via the SSE and converted into the electrical signal by the inverse SHE in IrMn, the effect is only weakly temperature dependent. From this it might look surprising that  the antidamping-like torque generated by the SHE is sizable only in the low-temperature antiferromagnetic phase of IrMn while at higher temperatures it is suppressed and fluctuates around zero, as summarized in Fig.~\ref{f4}(a).  Our trilayer is, however, not inversion symmetric and the broken reciprocity between the SSE and the current-induced torque  at higher temperatures can be reconciled as follows: 

In the SSE the spin-current is generated in the CoFeB and converted via the inverse SHE to the charge signal primarily in the IrMn layer underneath. In the current-induced torque, on the other hand, the spin-current is generated by the SHE in the Ta/IrMn bilayer. Since the spin Hall angles are opposite in Ta and IrMn, the contributions  from the two layers tend to compensate, resulting, at high temperatures, in a small net torque in CoFeB fluctuating around zero. 

At low temperatures when IrMn is antiferromagnetic the behavior changes. The torque in CoFeB becomes sizable and, considering its sign, it is dominated by the IrMn SHE.  The effect of the SHE spin-current from Ta on CoFeB moments is diminished. From experiments in the control Ta/CoFeB sample we observed that the SHE in Ta is temperature independent over the whole studied temperature range. Combining all these indications we arrive at a conclusion that the SHE spin-current from Ta is absorbed at low temperatures by the AFM moments in IrMn. 

Note the for the CoFeB and IrMn films strongly coupled at low temperatures the above scenario should be extended to account for the CoFeB/IrMn bilayer behaving as one joint magnetic medium. The calibration from first-harmonic resistances in external magnetic field then describes effective magneto-transport coefficients of the coupled bilayer. Similarly, $H_{AD}$  inferred from the second-harmonic resistances represents an effective current-induced field acting on the CoFeB/IrMn bilayer which can have contributions from torques generated at the Ta/IrMn and IrMn/CoFeB interfaces. The same applies to the spin Hall angle obtained form $H_{AD}$ and normalized to the current flowing through IrMn which parametrizes the entire NM/AFM/FM trilayer rather than the SHE of IrMn alone.

In either of the detailed versions of the above scenario, the current-induced  torque in the AFM plays a central role. To independently identify this effect we performed measurements in the Ta/IrMn  structure comprising no FM layer.  Since the IrMn moments in the low-temperature AFM phase cannot be manipulated by the applied magnetic field we use a procedure in which magnetic field is applied at different angles while cooling the Ta/IrMn sample from room-temperature. In earlier studies of similar thin IrMn films embedded in tunnel junctions, the field-cooling procedure enabled us to define states showing distinct magnetoresistance signals at low temperatures \cite{Petti2013}. These distinct states of the AFM IrMn then remained insensitive to the applied magnetic field when staying sufficiently below the N\'eel temperature \cite{Petti2013}. 

Analogous behavior of our Hall bar device is illustrated  in  Fig.~\ref{f4}(b). Here the sample was first cooled from 300~K to 5~K in an applied in-plane magnetic field of 2~T and then the experiment was repeated with field-cooling in a 2~T out-of-plane field. A clear onset of a non-zero difference $\Delta R_{xy}^{1\omega}$  between the two field-cool measurements is observed in the first harmonic resistance signal at $\sim 100$~K. This is consistent with the expected transition into the AFM phase in the thin IrMn film. It coincides with the observed enhancement of $H_{AD}$ at low temperatures in the Ta/IrMn/CoFeB stack which we also associated with the paramagnetic to AFM transition in IrMn (see Fig.~\ref{f4}(a)). Remarkably, the onset of $\Delta R_{xy}^{1\omega}$ in the Ta/IrMn bilayer is accompanied by the onset of the second harmonic $\Delta R_{xy}^{2\omega}$, suggesting that these signals are of magnetic origin and that the magnetic moments are torqued in the AFM by the applied current. 

To confirm that the signals in  Fig.~\ref{f4}(b) are due to the AFM moments  we show in Fig.~\ref{f4}(c) the second harmonic signal while sweeping the external magnetic field between $\pm2$~T and keeping the temperature at 5K. Similar to the first harmonic magnetoresistance, the second harmonic becomes completely insensitive to the strong magnetic fields in the low-temperature AFM phase of IrMn. Low-temperature magnetization data plotted in  Fig.~\ref{f4}(d),  showing only a diamagnetic signal of the substrate, provide an additional confirmation of the low-temperature AFM phase of IrMn and of the absence of a FM component in the stack.

We conclude by recalling that  theoretical studies have predicted relativistic current-induced torques in bulk AFMs as well as at interfaces \cite{Zelezny2014}. The bulk effects require specific symmetries of the crystal and magnetic structures of the AFM material and have been recently identified in current-induced switching experiments in CuMnAs \cite{Wadley2015}. The AFM torques acting at interfaces are expected to be more generic regarding the symmetries of the AFM unit cell. Our AFM-torque interpretation of the non-ohmic transport signals in the Ta/IrMn bilayer  is consistent with these theory predictions for AFM interfaces. Independently,  the SHE spin-current from Ta absorbed via the AFM torque in IrMn consistently fits into the overall picture obtained from  the complementary measurements in the Ta/IrMn, Ta/IrMn/CoFeB and Ta/CoFeB structures. Our results open the prospect for using current induced torques in transition metal multilayers to manipulate moments in AFMs.


We acknowledge  support from the EU European Research Council (ERC) advanced grant no. 268066, from the Marie Curie Initial Training Network grant no. 316657, from the Ministry of Education of the Czech Republic grant no. LM2011026, from the Grant Agency of the Czech Republic grant no. 14-37427G, and from the Austrian Science Fund (FWF): J3523-N27.

%

\end{document}


\title{Supplementary Material: \\ Current induced torques in structures with ultra-thin IrMn antiferromagnet}

\author{H.~Reichlov\'a}
\affiliation{Institute of Physics ASCR, v.v.i., Cukrovarnick\'a 10, 162 53
Praha 6, Czech Republic}
\affiliation{Faculty of Mathematics and Physics, Charles University in Prague,
Ke Karlovu 3, 121 16 Prague 2, Czech Republic}
\author{D.~Kriegner}
\affiliation{Faculty of Mathematics and Physics, Charles University in Prague,
Ke Karlovu 3, 121 16 Prague 2, Czech Republic}
\author{V.~Hol\'y}
\affiliation{Faculty of Mathematics and Physics, Charles University in Prague,
Ke Karlovu 3, 121 16 Prague 2, Czech Republic}
\author{K.~Olejn\'{\i}k}
\affiliation{Institute of Physics ASCR, v.v.i., Cukrovarnick\'a 10, 162 53
Praha 6, Czech Republic} 
\author{V.~Nov\'ak}
\affiliation{Institute of Physics ASCR, v.v.i., Cukrovarnick\'a 10, 162 53
Praha 6, Czech Republic} 
\author{M.~Yamada}
\affiliation{Hitachi Ltd., Central Research Laboratory, 1-280 Higashi-koigakubo, Kokubunju-shi, Tokyo 185-8601, Japan} 
\author{K.~Miura}
\affiliation{Hitachi Ltd., Central Research Laboratory, 1-280 Higashi-koigakubo, Kokubunju-shi, Tokyo 185-8601, Japan} 
\author{S.~Ogawa}
\affiliation{Hitachi Ltd., Central Research Laboratory, 1-280 Higashi-koigakubo, Kokubunju-shi, Tokyo 185-8601, Japan} 
\author{H.~Takahashi}
\affiliation{Hitachi Ltd., Central Research Laboratory, 1-280 Higashi-koigakubo, Kokubunju-shi, Tokyo 185-8601, Japan} 
\author{T.~Jungwirth}
\affiliation{Institute of Physics ASCR, v.v.i., Cukrovarnick\'a 10, 162 53
Praha 6, Czech Republic} 
\affiliation{School of Physics and
Astronomy, University of Nottingham, Nottingham NG7 2RD, United Kingdom}
\author{J.~Wunderlich}
\affiliation{Institute of Physics ASCR, v.v.i., Cukrovarnick\'a 10, 162 53 Praha 6, Czech Republic}
\affiliation{Hitachi Cambridge Laboratory, Cambridge CB3 0HE, United Kingdom}

\maketitle

\section{Sample growth and fabrication}
Ta, IrMn, and CoFeB layers were deposited by UHV RF magnetron sputtering on a thermally oxidized Si substrate (700~nm SiO$_2$ on (001) Si) at a base pressure of 10$^{-9}$~Torr and in a magnetic field of 5~mT. After growth, the wafers were annealed at 350$^\circ$C for 1~hour in a 10$^{-6}$~Torr vacuum in a magnetic field of 0.6~T. Annealing of CoFeB (20:60:20) ferromagnetic layer leads to crystallization of the layer and provides better magnetic properties \cite{Hayakawa2005,Hayakawa2005a}. The five distinct layers (not alloyed) were confirmed by X-Ray reflectivity (see next section). For transport measurements Hall bars were defined in the HSQ negative resist by e-beam lithography. The contacts were defined by an aluminum mask and the stack of layers was etched everywhere else by ion milling. The width of the bars is 2~$\mu$m and the distance between longitudinal contacts is 7~$\mu$m. Several samples were patterned from the wafer, all showing similar results.

\section{X-Ray characterization}
\begin{figure}[h!]
\hspace*{0cm}\epsfig{width=.5\columnwidth,angle=0,file=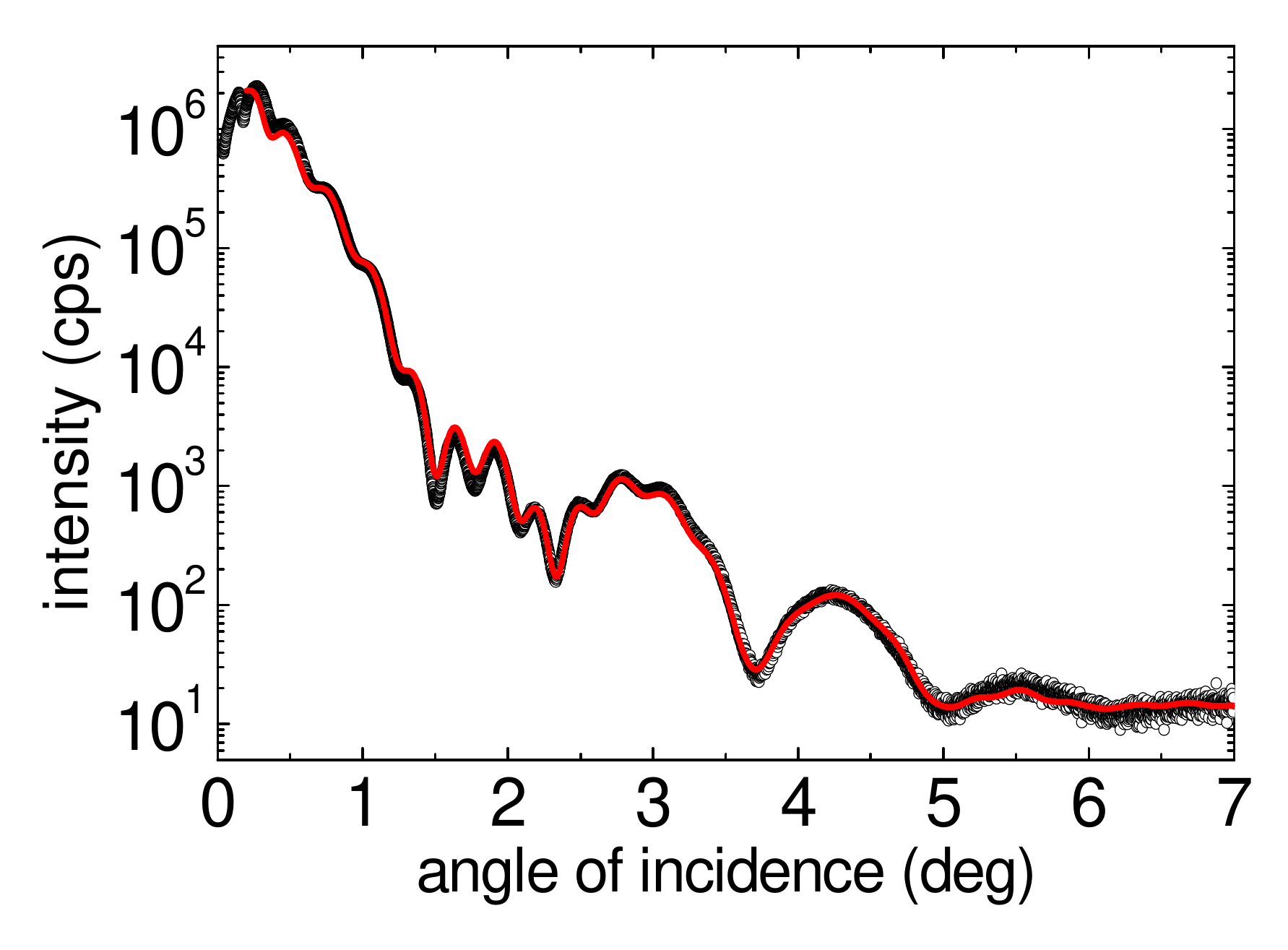}
\vspace*{-0.5cm}
\caption
{X-ray reflectivity measurement (black dots) and simulation (red line).}
\label{Sf1}
\end{figure}

X-ray reflectivity (XRR) measurements were performed using a laboratory diffractometer with Cu radiation. Results for the Ta/IrMn/CoFeB sample are shown in Fig.~\ref{Sf1}. The layer thicknesses infirmed from XRR are: Ta($2.2\pm0.2$~nm)/IrMn($0.6\pm0.3$~nm)/CoFeB($0.9\pm0.2$~nm)/MgO($1.4\pm0.2$~nm) capped with 10~nm of Al$_2$O$_3$. The root mean square roughness of all interfaces is $0.15-0.3$~nm. XRR simulations  shown in Fig.~\ref{Sf1} are performed using the Parratt formalism \cite{VaclavXRay}. Omitting any of the five distinct layers in the simulation would not allow to fit the measured data.

\section{Transport measurements in magnetic field sweeps} 
\begin{figure}[h!]
\hspace*{0cm}\epsfig{width=.5\columnwidth,angle=0,file=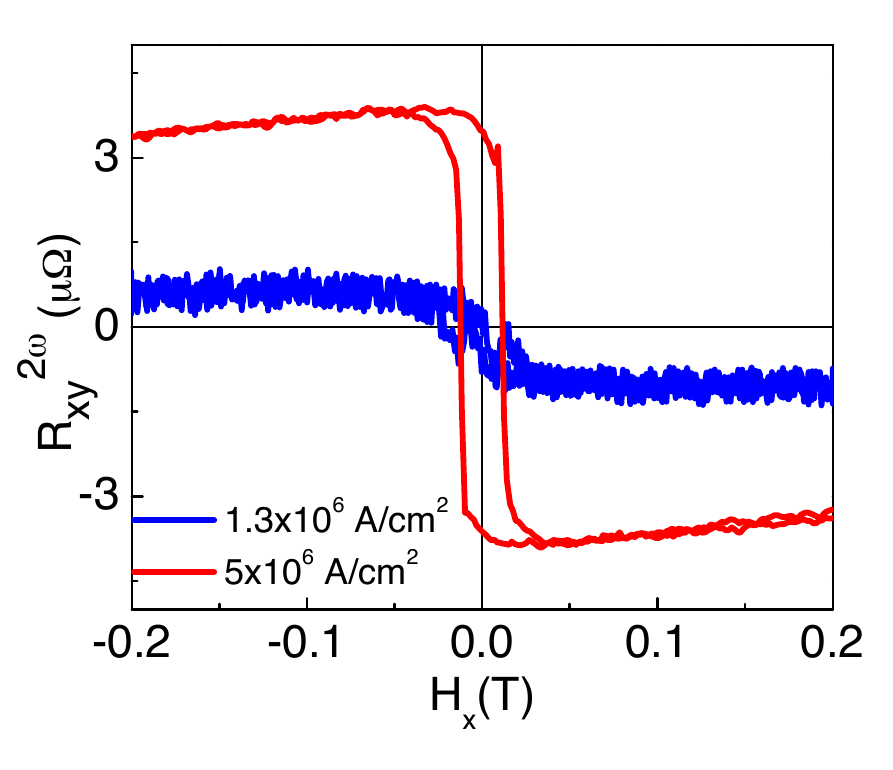}
\vspace*{-0.5cm}
\caption
{$R_{xy}^{2\omega}$ measured when magnetic field is swept along the current direction for two current densities.}
\label{Sf2}
\end{figure}

In order to confirm the magnetic origin of  the second-harmonic resistance signals, we performed transport measurements while sweeping the magnetic field in the plane of the sample (the experimental geometry corresponding to the SQUID measurement in Fig. 1(a) of the main text). The  data are shown in Fig.~\ref{Sf2}.  $R_{xy}^{2\omega}$  increases with current density and also reflects the hysteretic magnetization switching. The coercive field is broader than what is observed by SQUID which we attribute to different domain structure and pinning in the two exeriments. Note that the magnetotransport data are measured on a narrow Hall bar while for magnetometry measurements we used an unpatterned piece  of the wafer. Since the exchange bias is weak (only a few Oersted), it is not detected  in our cryostat for magneto-transport measurements where a precision of a few Oersted is not achievable.

\section{Temperature dependence of the first harmonic resistance signals}
\begin{figure}[H]
\hspace*{0cm}\epsfig{width=1\columnwidth,angle=0,file=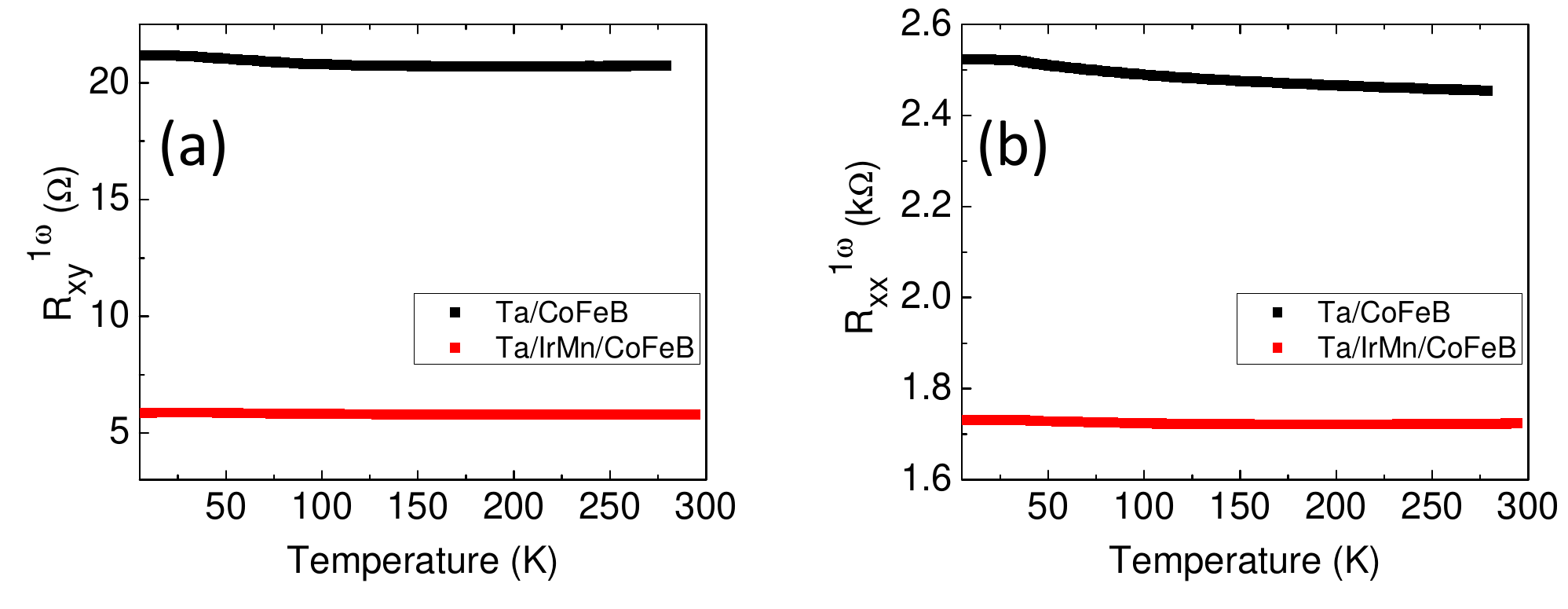}
\vspace*{-0.5cm}
\caption
{Temperature dependence of the $R_{xy}^{1\omega}$ (a) $R_{xx}^{1\omega}$ (b) signals.}
\label{Sf11}
\end{figure}
Within the studied temperature range, the resistance of our metallic stacks  depends only weakly on the base temperature,  as shown in Fig.~\ref{Sf11}. The  thermal gradients due to Joule heating across the metal multilayers should, therefore, be also only weakly temperature dependent. 

\begin{figure} [h!]
\centering
\includegraphics[scale=.7]{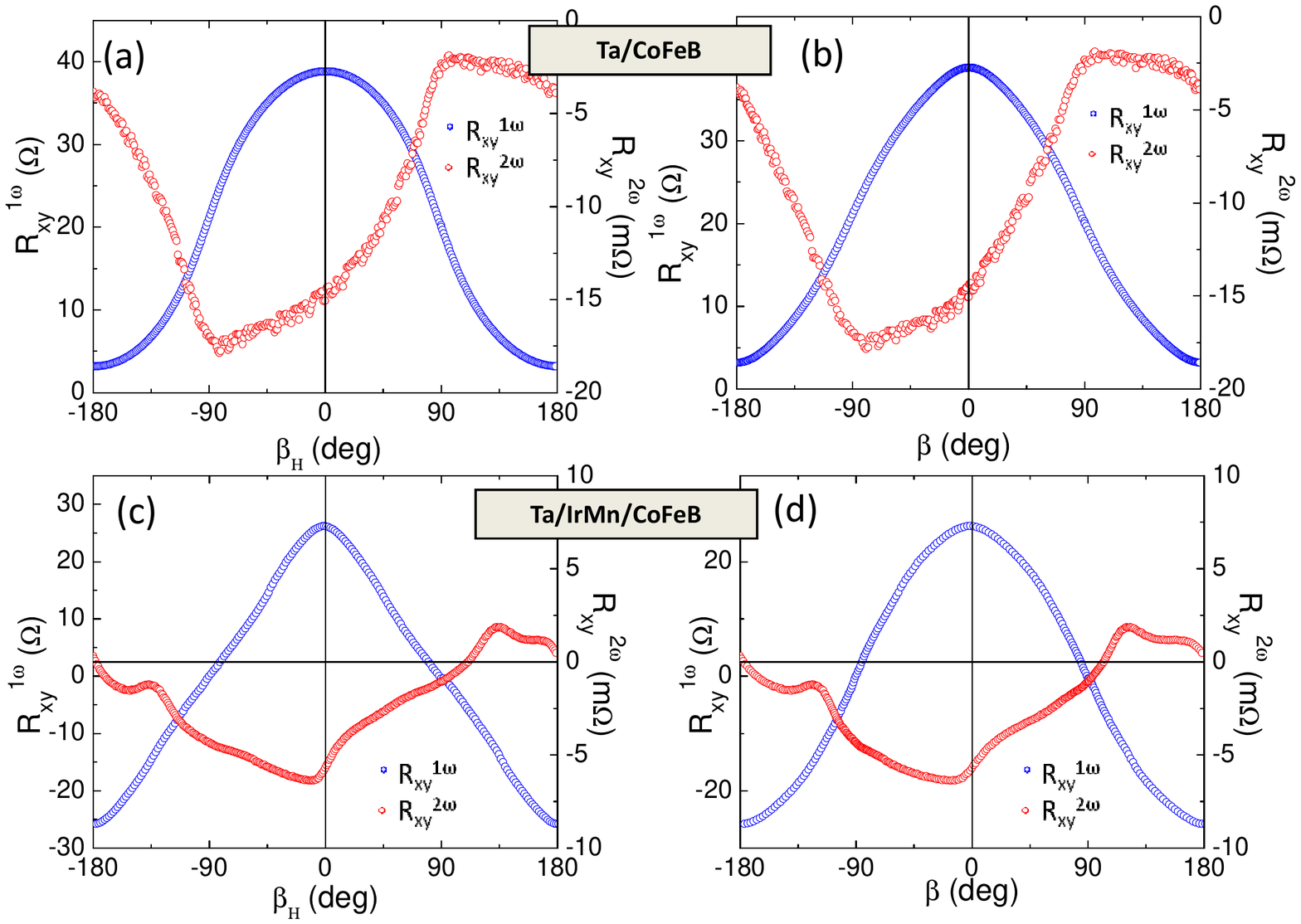} 
\caption{\label{Sfig10} $R_{xy}^{1\omega}$ (red symbols) and $R_{xy}^{2\omega}$ (blue symbols) as a function of the magnetic field angle $\beta_H$ for field rotation in the x-z plane (left) and as a function of the corresponding magnetization angle $\beta$ (right).}
\end{figure}

\section{Magnetic field vs. magnetization in out-of-plane rotation measurements}
In the  Ta/CoFeB bilayer, the FM CoFeB has an out-of-plane easy-axis with a corresponding anisotropy field of 0.5~T. In the Ta/IrMn/CoFeB trilayer, the FM CoFeB is in-plane magnetized and the anisotropy field is -0.6~T. Because of these anisotropies, the angle of magnetization does not follow the magnetic field angle in the out-of-plane rotation measurements in a 1~T field.  To correct our analysis for these deviations we recalculate the magnetic field angle  to the corresponding magnetization angles as illustrated in Fig.~\ref{Sfig10}. 

\newpage

\section{Analysis of the second-harmonic transport signals}

\begin{figure} [h!]
\centering
\includegraphics[scale=.6]{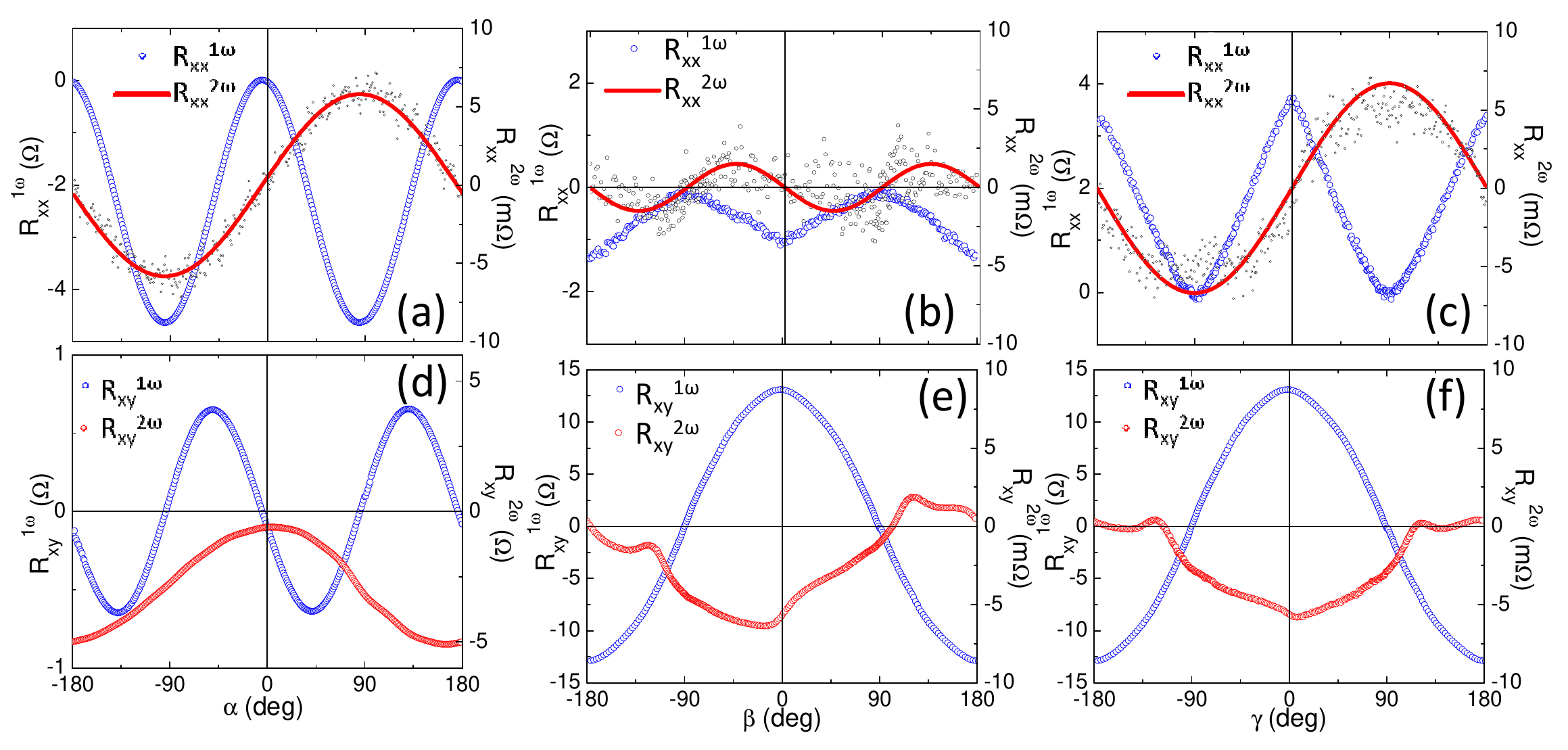} 
\caption{\label{Sfig9} Sample Ta/IrMn/CoFeB. First harmonics (blue symbols) and second harmonics (red symbols) when magnetization rotates in all three planes measured on longitudinal (panels (a)-(c)) and transverse (panels (d)-(f)) contacts.  Data are measured  at 5K at $j=3.5~A/cm^2$.}
\end{figure}

\begin{figure} [h!]
\centering
\includegraphics[scale=.6]{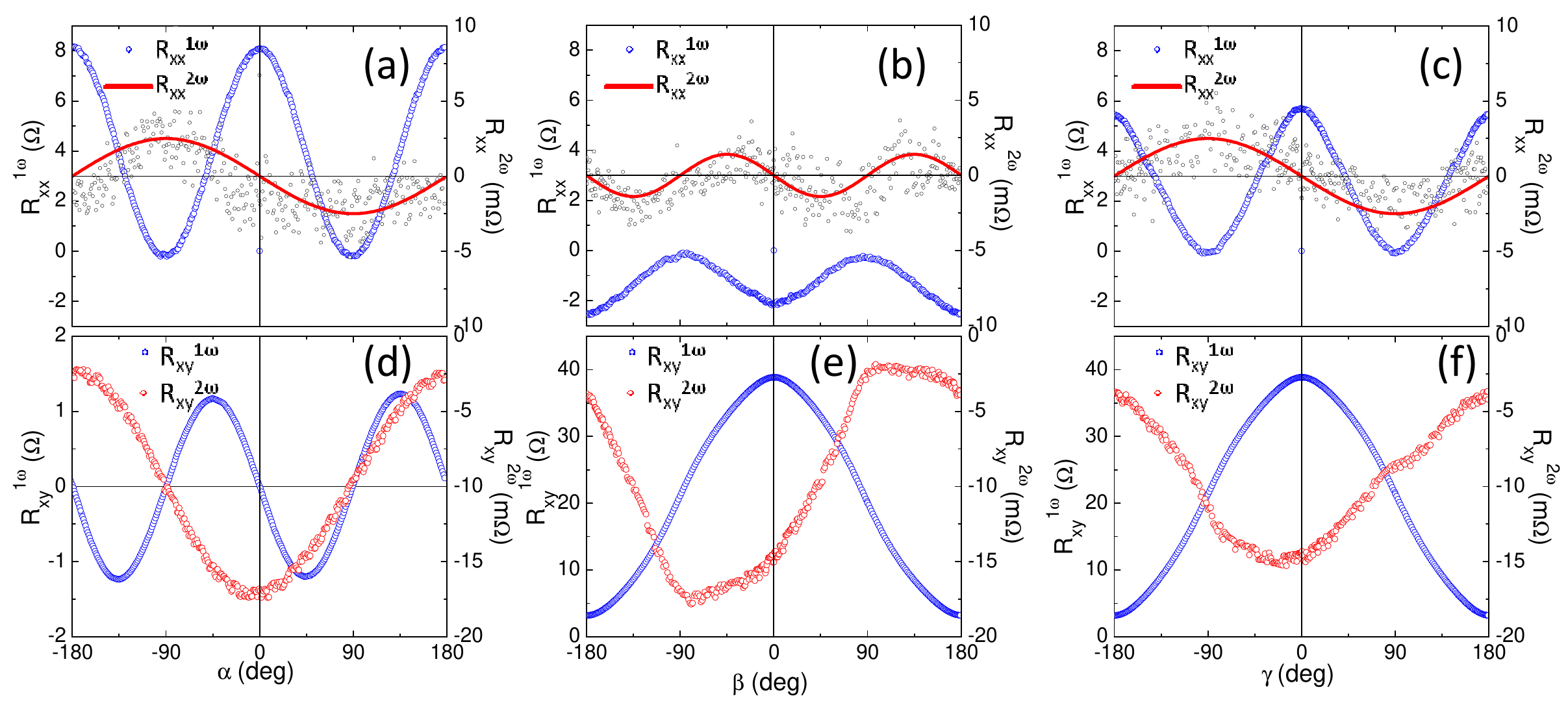} 
\caption{\label{Sfig14} Sample Ta/CoFeB. First harmonics (blue symbols) and second harmonics (red symbols) when magnetization rotates in all three planes measured on longitudinal (panels (a)-(c)) and transverse (panels (d)-(f)) contacts. Data measured  at 5K at $j=3.6~A/cm^2$.}
\end{figure}

Examples of the measured transport signals in the Ta/IrMn/CoFeB and Ta/CoFeB samples are shown in Figs.~\ref{Sfig9} and \ref{Sfig14}.

\newpage

\begin{figure}[h!]
\hspace*{0cm}\epsfig{width=0.3\columnwidth,angle=0,file=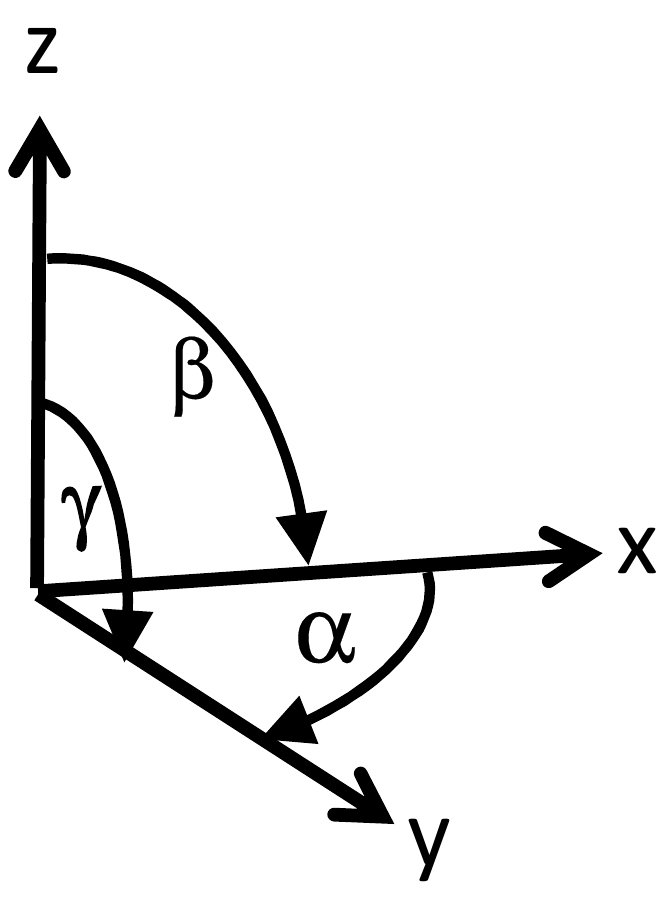}
\vspace*{-0.5cm}
\caption
{Definitions of magnetization rotation angles. x-y is the sample plane and current is applied along the x-direction.}
\label{Sfig8}
\end{figure}

We now describe our analysis of terms contributing to the second harmonic transport signal as a function of the rotation angles (Fig.~\ref{Sfig8}) of the equilibrium magnetization vector.  We consider the following magnetotransport coefficients that are sensitive to the magnetization vector: the magnetization dependence of the off-diagonal $R_{xy}$ component is due to the Anomalous Hall effect (AHE) and the transverse anisotropic magnetoresistance (tAMR) (also called the planar Hall effect); the diagonal  $R_{xx}$ component can depend on the magnetization via the non-crystalline anisotropic magnetoresistance (ncAMR) determined by the angle between the magnetization and current, and the crystalline anisotropic magnetoresistance (cAMR) depending on the angle of the magnetization from the out-of-plane axis. 

The AHE has an extremum when magnetization is perpendicular to the sample plane. The tAMR has  a maximum  at a $45^\circ$ angle between the in-plane projection of the magnetization and current while the ncAMR in our samples has a maximum  when magnetization and current are collinear. The cAMR has  a maximum for the out-of-plane magnetization direction.  

The above terms can contribute to the second-harmonic signals when magnetization is tilted from the equilibrium position by the current-induced torques. The effective  field ${\bf H}_{AD}\sim {\bf m}\times{\bf \hat{y}}$ corresponds to the antidamping-like torque while the effective field  ${\bf H}_{F}\sim{\bf \hat{y}}$ corresponds to the field-like torque. Microscopically $H_{AD}$ can originate, e.g., from the SHE while $H_{F}$ from, e.g., the Oersted or Rashba fields.

An out-of-plane thermal gradient $\Delta T_{oop}$ can contribute to the magnetization dependence of the second-harmonic transport signals without having induced any tilt of the magnetization out of the equilibrium direction.   This can be due to the Anomalous Nernst effect (ANE) or the spin Seebeck effect (SSE). Both effects have the same symmetry with the maximum contribution when the magnetization is in-plane and perpendicular to the axis connecting the respective voltage probes in $R_{xx}$ and $R_{xy}$ measurements. Below we show that the sign of this effect is opposite in the Ta/IrMn/CoFeB and Ta/CoFeB samples which implies that the effect is due to the SSE. An in-plane thermal gradient $\Delta T_{ip}$ along the applied current can contribute to the second-harmonic $R_{xy}$ signal through the ANE. The maximum contribution in this case is for the out-of-plane magnetization.

All the above contributions and their respective angular dependencies are summarized in Tabs.~\ref{2wcontributionsRxx} and \ref{2wcontributionsRxy}.

\begin{table}[h!]
   \centering
			    \begin{tabular}{ | l || l | l | l | }
    \hline
   $R_{xx}^{2\omega}$ & $\alpha$& $\beta$ & $\gamma$  \\  \hline \hline
   $R_{\text{ncAMR}}$ & $r_\text{ncAMR}H_{F}\sin(2\alpha)\cos\alpha$ & $-\bar{r}_\text{ncAMR}H_{AD}\sin(2\beta)$ &  $0$  \\ \hline
   $R_{\text{cAMR}}$ & $0$ & $-\bar{r}_\text{cAMR}H_{AD}\sin(2\beta)$ & $2\bar{r}_\text{cAMR}H_{F}\sin(2\gamma)\cos\gamma $\\  \hline
   thermal & $r_\text{SSE}\Delta T_{oop} \sin\alpha$ &  $0$ &  $r_\text{SSE}\Delta T_{oop}\sin\gamma$  \\ \hline
   				\end{tabular}
\caption{\label{2wcontributionsRxx} Angular dependence of the contributions to the second harmonic signal $R_{xx}^{2\omega}$. $r_i$ are the coefficients for the various terms discussed in the text.}
\end{table}
\begin{table}[h!]
	
        \centering
			    \begin{tabular}{ | l || l | l | l |}
    \hline
  $R_{xy}^{2\omega}$  & $\alpha$& $\beta$ & $\gamma$  \\  \hline \hline
   $R_{\text{AHE}}$ &  $r_\text{AHE}H_{AD}\cos\alpha $ &  $\bar{r}_\text{AHE}H_{AD}\sin\beta $ &  $-\bar{r}_\text{AHE}H_{F}\sin\gamma\cos\gamma $  \\  \hline
   $R_{\text{tAMR}}$ &  $r_\text{ncAMR}2H_{F}\cos(2\alpha)\cos\alpha $ & $\bar{r}_\text{ncAMR}2H_{F}\operatorname{sgn}(\sin\beta) $&$\bar{r}_\text{ncAMR}2H_{AD}\operatorname{sgn}(\sin\gamma)\cos\gamma$ \\ 
 \hline
   thermal &  $ r_\text{SSE}\Delta T_{oop}\cos\alpha$ & $ r_\text{SSE}\Delta T_{oop}\sin\beta$ &  $r_\text{ANE}\Delta T_{ip}\cos\gamma$ \\ 
   			 &  & $ + r_\text{ANE}\Delta T_{ip}\cos\beta$ &    \\ \hline
   				\end{tabular}
\caption{\label{2wcontributionsRxy} Same as Tab.~\protect\ref{2wcontributionsRxx} for  $R_{xy}^{2\omega}$.}
\end{table}

First harmonic measurements of the AHE and AMR with magnetization rotated in the three planes by an external magnetic field are used to calibrate the corresponding coefficients $r_i$ and $\bar{r}_i$ in Tabs.~\ref{2wcontributionsRxx} and \ref{2wcontributionsRxy}. Note that the total effective field aligning the equilibrium magnetization along the angles  $\alpha$, $\beta$, or $\gamma$ comprises the externally applied magnetic field and the anisotropy field. The anisotropy field for the out-of-plane rotations  is sizable compared to the externally applied magnetic field. The total equilibrium effective field is absorbed in the coefficients  $r_i$ and $\bar{r}_i$ and, therefore, they are different for the in-plane and out-of-plane rotations. 

The $\alpha$ and $\gamma$-rotation $R_{xx}^{2\omega}$ signals are dominated in both the Ta/IrMn/CoFeB and Ta/CoFeB samples by $\sin\alpha$ and $\sin\gamma$ terms, respectively. As seen from Tab.~\ref{2wcontributionsRxx}, these terms are due to the  SSE. We can therefore use these  $R_{xx}^{2\omega}$ signals to calibrate the amplitude $r_\text{SSE}\Delta T_{oop}$ of the out-of-plane thermal gradient effect.  The current induced effective fields $H_{AD}$ and  $H_{F}$ are then inferred from the measured second-harmonic transport signals by decomposing the data into the individual angle-dependent terms and by using the above calibrations.


\section{Analysis of the Second Harmonic Signal}

\begin{figure} [h!]
\centering
\includegraphics[scale=0.8]{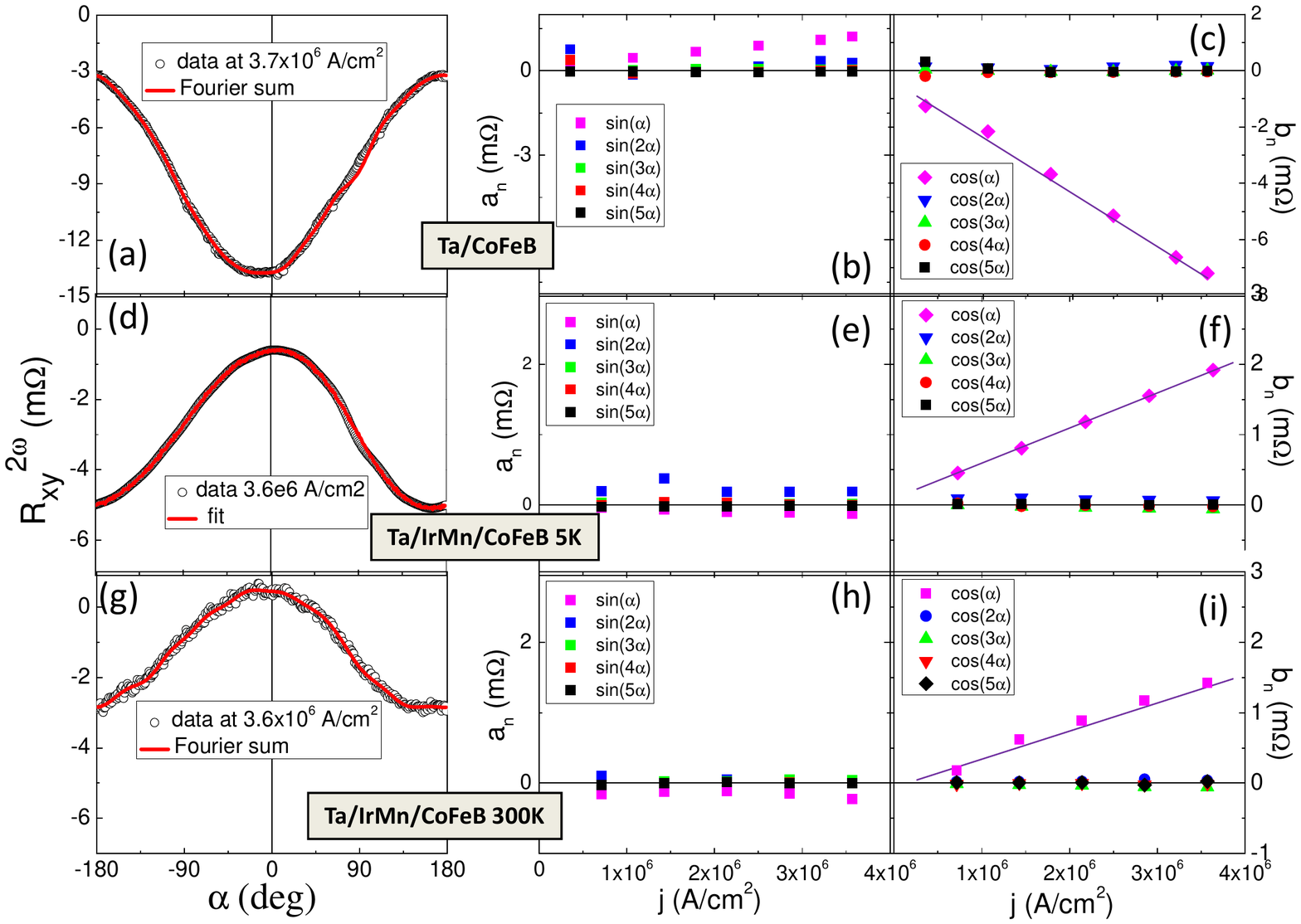} 
\caption{\label{Sfig4} Left: The $R_{xy}^{2\omega}(\alpha)$ signal (black circles) when magnetization is rotated in-plane and the corresponding Fourier sum fitting  (red line). Right: Inferred Fourier coefficients as a function of the current density.}
\end{figure}

\protect\newpage

\begin{figure}[h!]
\centering
\includegraphics[scale=0.8]{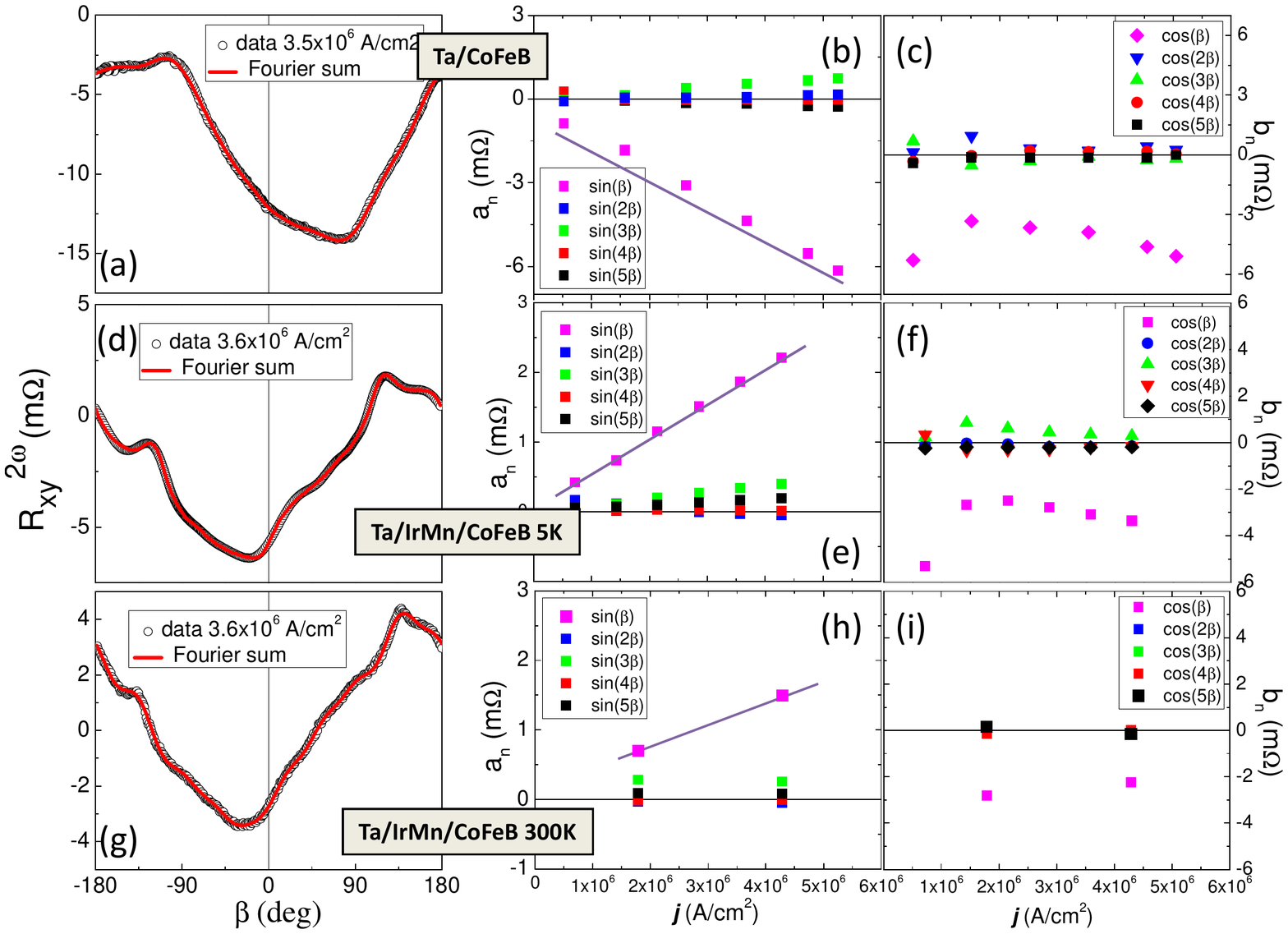} 
\caption{\label{Sfig12} Same as Fig.~\protect\ref{Sfig4} for the out-of-plane $\beta$-rotation of the magnetization.}
\end{figure}

\protect\newpage

\begin{figure}[h!]
\centering
\includegraphics[scale=0.8]{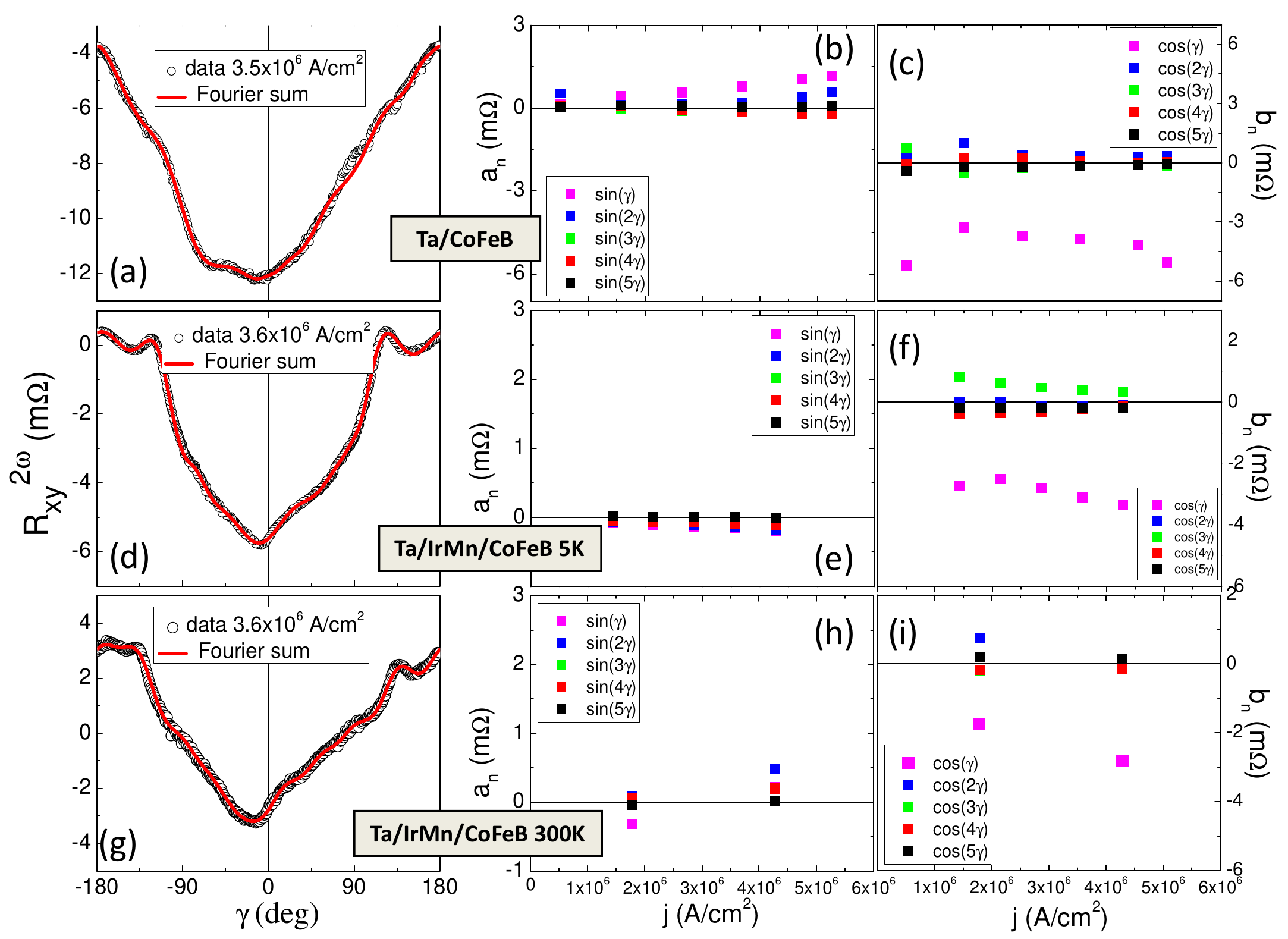} 
\caption{\label{Sfig13} Same as Fig.~\protect\ref{Sfig4} for the out-of-plane $\gamma$-rotation of the magnetization.}
\end{figure}


Since the $\beta$-dependent $R_{xx}^{2\omega}$ has the expected $\sin2\beta$ dependence we can use it directly to obtain $H_{AD}$. The results are consistent with $H_{AD}$ inferred from $R_{xy}^{2\omega}$. The $R_{xy}^{2\omega}$ signals are less noisy and thus allow us for a more precise determination of the current induced effective fields. However, their angular dependencies are more complex. To analyze the $R_{xy}^{2\omega}$ data, we decomposed the measured signals for all three magnetization rotation angles   into harmonic functions of the respective magnetization angles with  coefficients $a_n$ and $b_n$ defined as
\begin{align}
a_n=\frac{1}{\pi} \int_0^{2\pi} f(x)\sin(nx) dx;~
b_n=\frac{1}{\pi} \int_0^{2\pi} f(x)\cos(nx) dx \,.
\end{align}
Here $f(x)$ is our measured signal and $x$ represents the angle of rotation. The measured data are than described by the Fourier sum:
\begin{equation}
f(x)=\sum_{n=1}^N (a_n \sin(nx) + b_n \cos(nx))+ \text{const.}
\label{FourierSum}
\end{equation}

Results of the Fourier analysis are summarized in Figs.~\ref{Sfig4}, \ref{Sfig12}, and \ref{Sfig12} for the $\alpha$, $\beta$, and $\gamma$-rotation, respectively.  Consistent values of $H_{AD}$, quoted in the main text,  are obtained from the $\cos\alpha$ and $\sin\beta$ terms in $R_{xy}^{2\omega}$. As expected, the terms scale linearly with the applied current. All other terms, except for the $\cos\beta$ and $\cos\gamma$ components, are negligible in $R_{xy}^{2\omega}$  which implies that we do not observe a sizable $H_F$ field in our measurements.

The $\cos\beta$ and $\cos\gamma$ terms in $R_{xy}^{2\omega}$ do not scale with the applied electrical current and are, therefore, not related to the current induced torques. We attribute them to the in-plane thermal gradient effect. As expected for the ANE generated in the FM, the terms have the same sign in the Ta/IrMn/CoFeB and Ta/CoFeB samples. The different scaling with applied current, compared to the out-of-plane thermal gradient effect, might reflect the different Joule heating characteristics in the geometry when electrons drift along the thermal gradient. The drifting electrons between the warmer and colder longitudinal contacts will tend to equilibrate the system, resulting in the weak net dependence of the corresponding $R_{xy}^{2\omega}$ component on current.

%